\documentclass[noshowpacs,amsmath,
twocolumn,
superscriptaddress,
8pt,
aps,prb]{revtex4-1}
\usepackage{setspace}
\usepackage{amsmath}
\usepackage{graphicx}
\usepackage[nearskip,margin = 0pt]{subfig}

\usepackage{verbatim}
\usepackage{amsfonts}
\usepackage{amssymb}
\usepackage{epstopdf} 
\usepackage{xcolor}
\usepackage{verbatim}
\DeclareGraphicsExtensions{.pdf,.eps,.png,.jpg,.mps} 
\begin{document}
\title{Vernier spectrometer using counter-propagating soliton microcombs}

\author{Qi-Fan Yang$^{1,\ast}$, Boqiang Shen$^{1,\ast}$, Heming Wang$^{1,\ast}$, Minh Tran$^{2}$, Zhewei Zhang$^{1}$, Ki Youl Yang$^{1}$, Lue Wu$^{1}$, Chengying Bao$^{1}$, John Bowers$^{2}$, Amnon Yariv$^{1}$ and Kerry Vahala$^{1,\dagger}$\\
$^{1}$T. J. Watson Laboratory of Applied Physics, California Institute of Technology, Pasadena, California 91125, USA.\\
$^{2}$University of California, Santa Barbara, Department of Electrical and Computer Engineering, Santa Barbara, CA 93106, USA.\\
$^{\ast}$These authors contributed equally to this work.\\
$^{\dagger}$Corresponding author: vahala@caltech.edu}

\date{\today}

\maketitle


\noindent {\bf Acquisition of laser frequency with high resolution under continuous and abrupt tuning conditions is important for sensing, spectroscopy and communications.  Here, a single microresonator provides rapid and broad-band measurement of frequencies across the optical C-band with a relative frequency precision comparable to conventional dual frequency comb systems. Dual-locked counter-propagating solitons having slightly different repetition rates are used to implement a Vernier spectrometer.  Laser tuning rates as high as 10 THz/s, broadly step-tuned lasers, multi-line laser spectra and also molecular absorption lines are characterized using the device. Besides providing a considerable technical simplification through the dual-locked solitons and enhanced capability for measurement of arbitrarily tuned sources, this work reveals possibilities for chip-scale spectrometers that greatly exceed the performance of table-top grating and interferometer-based devices.}


Frequency-agile lasers are ubiquitous in sensing, spectroscopy and optical communications \cite{wilner2008communications,allen1998diode,choma2003sensitivity} and measurement of their optical frequency for tuning and control is traditionally performed by grating and interferometer-based spectrometers, but more recently these measurements can make use of optical frequency combs\cite{jones2000carrier,holzwarth2000optical,diddams2010evolving}.  Frequency combs provide a remarkably stable measurement grid against which optical signal frequencies can be determined subject to the ambiguity introduced by their equally spaced comb lines. The ambiguity can be resolved for continuously frequency swept signals by counting comb teeth \cite{del2009frequency} relative to a known comb tooth; and this method has enabled measurement of remarkably high chirp rates \cite{coddington2012characterizing}. However, signal sources can operate with abrupt frequency jumps so as to quickly access a new spectral region or for switching purposes, and this requires a different approach.  In this case, a second frequency comb with a different comb line spacing can provide a Vernier scale \cite{giorgetta2010fast} for comparison with the first comb to resolve the ambiguity under quite general tuning conditions \cite{ma2003new,peng2008optical,giorgetta2010fast}. This Vernier concept is also used in dual comb spectroscopy\cite{coddington2016dual,suh2016microresonator}, but in measuring active signals the method can be significantly enhanced to quickly identify signal frequencies through a signal correlation technique \cite{giorgetta2010fast}. The power of the Vernier-based method relies upon mapping of optical comb frequencies into a radio-frequency grid of frequencies, the precision of which is set by the relative line-by-line frequency stability of the two frequency combs. This stability can be guaranteed by self-referencing each comb using a common high-stability radio-frequency source or through optical locking of each comb to reference lasers whose relative stability is ensured by mutual locking to a common optical cavity. 

\begin{figure*}
\captionsetup{singlelinecheck=off, justification = RaggedRight}
\includegraphics[width=18cm]{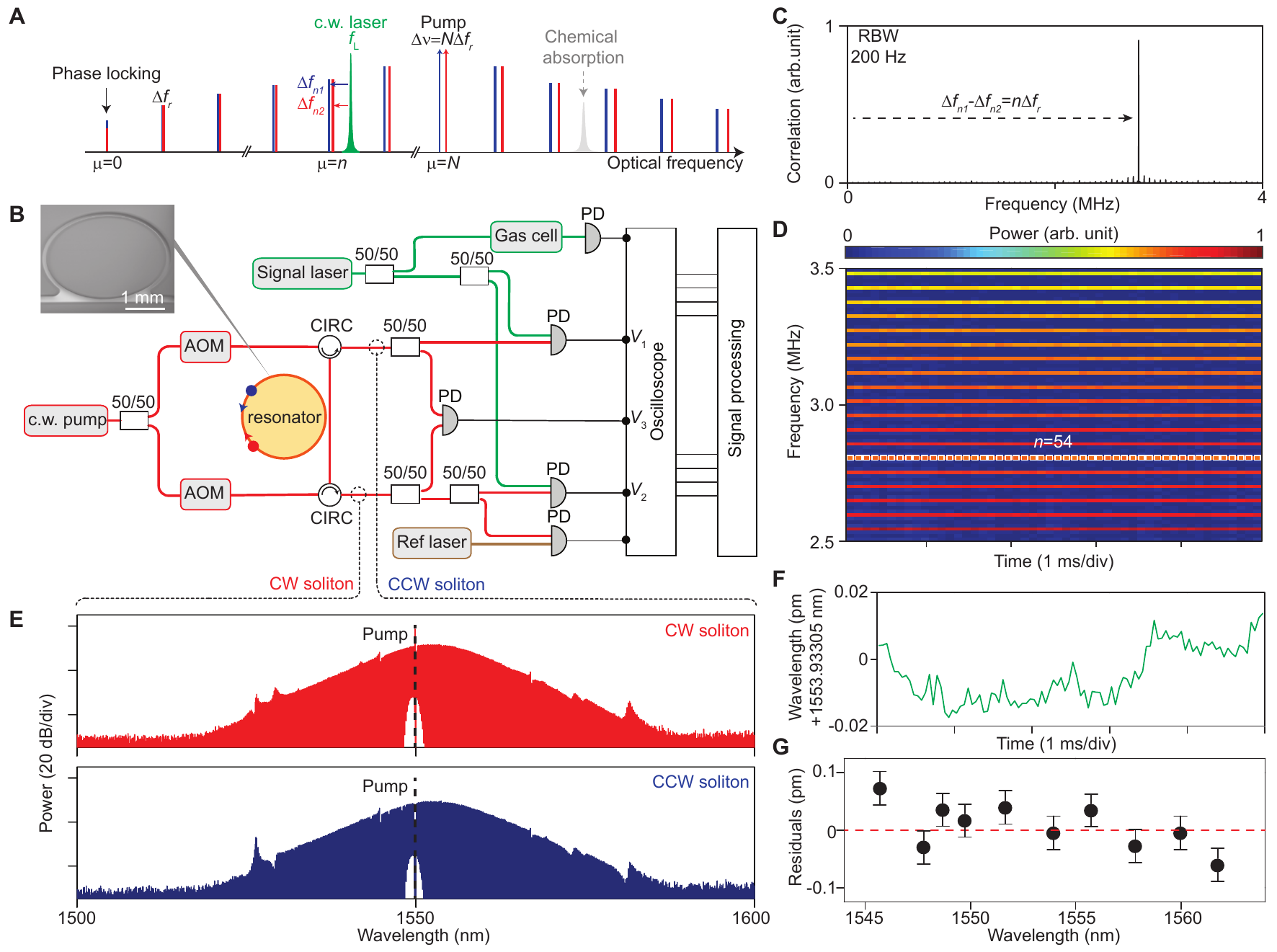}
 \caption{{\bf Spectrometer concept, experimental setup and static measurement.} ({\bf A}) Counter propagating soliton frequency combs (red and blue) feature repetition rates that differ by $\Delta f_r$. Their propagation in the resonator causes phase-locking at the comb line with index $\mu=0$. Also, the comb teeth separated by $\Delta \nu = N \Delta f_r$ at $\mu=N$ are derived from a single pump laser and therefore also are effectively locked. This dual locking of the vernier-like comb frequencies enables precise measurement of a laser (green) at frequency $f_L$ when combined with electrical correlation of the comb signals to determine $\mu = n$. Once calibrated, the tunable laser can resolve  chemical absorption lines (grey) with high precision. ({\bf B}) Experimental setup. AOM: acousto-optic modulator; CIRC: circulator; PD: photodetector. Supplement includes more detail. Inset: scanning electron microscope image of a silica resonator.  ({\bf C}) Typical measured spectrum of $V_1 V_2$ used to determine order $n$. For this spectrum: $\Delta f_{n1} - \Delta f_{n2}$ = 2.8052 MHz and $\Delta f_r=52$ kHz giving $n=54$.  ({\bf D}) The spectrograph of the dual soliton interferogram (pseudo color). Line spacing gives $\Delta f_r=52$ kHz. White squares correspond to the index $n=54$ in panel C. ({\bf E})  Optical spectra of counter-propagating solitons. Pumps are filtered and denoted by dashed lines.({\bf F}) Measured wavelength of an external cavity diode laser operated in steady state. ({\bf G}) Residual deviations between ECDL laser frequency measurement as given by  the MSS and a wavemeter.  Error bars give the systematic uncertainty as limited by the reference laser in panel B.}
\label{figure1}
\end{figure*}

Here, a broad-band, high-resolution Vernier soliton microcomb spectrometer is demonstrated using a single miniature comb device that generates two mutually-phase-locked combs. The principle of operation relies upon an optical phase locking effect observed in the generation of counter-propagating solitons within high-Q whispering gallery resonators \cite{yang2017counter}. Soliton generation in microcavities is being studied for miniaturization to the chip-scale of complete comb systems \cite{Kippenberg2018} and these so-called soliton microcombs have now been demonstrated in a wide range of microcavity systems \cite{herr2014temporal,yi2015soliton,brasch2016photonic,wang2016intracavity,joshi2016thermally,gong2018high}. In the counter-propagating soliton system, it is found that the clockwise (cw) and counter-clockwise (ccw) comb frequencies can be readily phase locked with distinct repetition rates that are also locked. This mutual double-locking creates line-by-line relative frequency stability for the underlying microcomb spectra that is more characteristic of fully self-referenced dual comb systems. The resulting Vernier of comb frequencies in the optical domain maps to an exceptionally stable radio frequency grid. Application of the signal correlation method \cite{giorgetta2010fast} to this system, then enables a microresonator soliton spectrometer (MSS) for rapid and high accuracy measurement of frequency.  

To establish its performance and for comparison with dual fiber-mode-locked-laser spectrometers \cite{giorgetta2010fast} the MSS is applied to measure a 10 THz/s laser frequency chirping rate, step tuning of a laser, as well as acquisition of high-resolution molecular vibronic spectra over the optical C-band.
Moreover, a method for signal frequency extraction is developed that uses the high relative stability of the cw and ccw combs to unambiguously determine frequencies in complex spectra containing 100s of frequencies.

\begin{figure*}
\captionsetup{singlelinecheck=off, justification = RaggedRight}
\begin{center}
\includegraphics{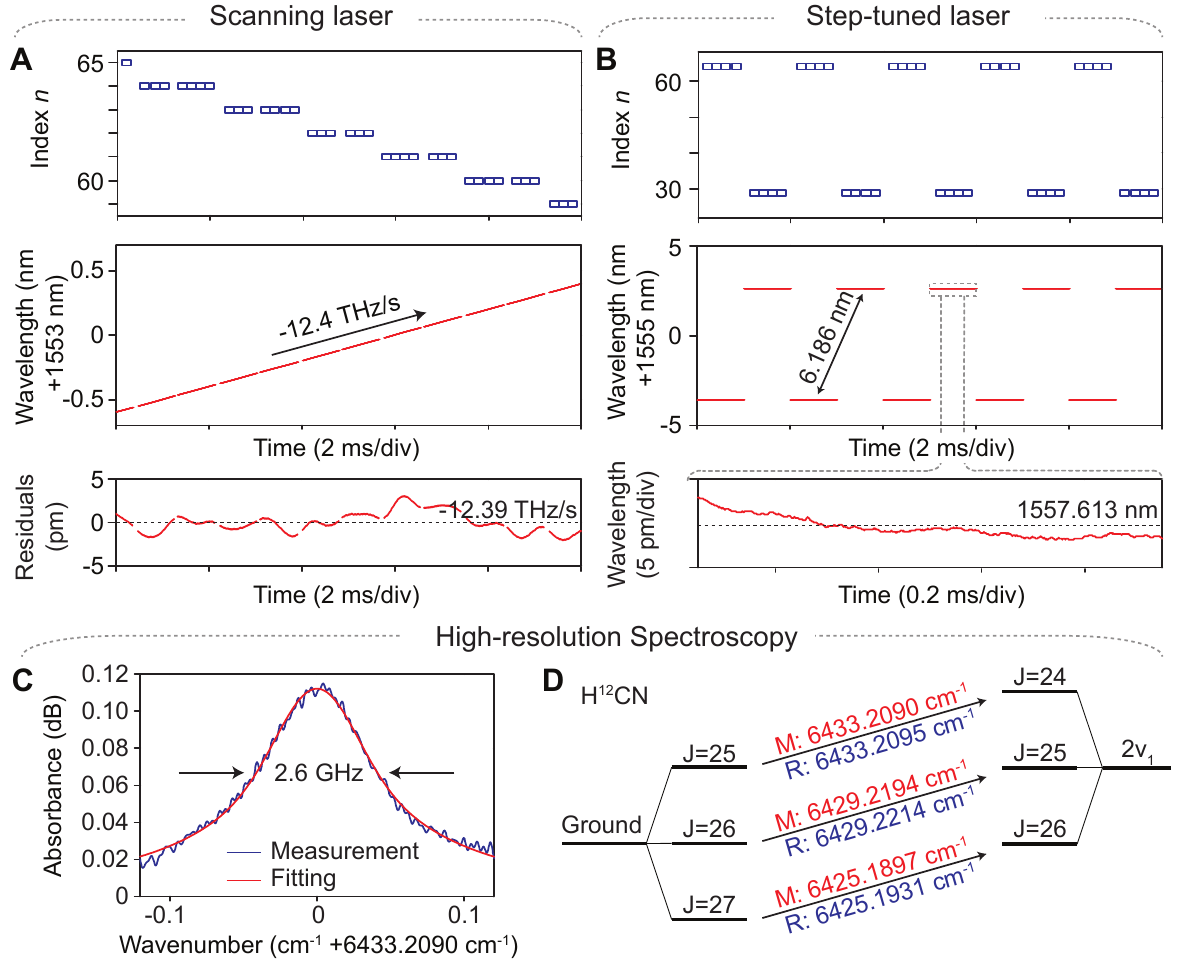}
\caption{{\bf Laser tuning and spectroscopy measurements.} ({\bf A}) Measurement of a rapidly tuning laser showing index $n$ (upper), instantaneous frequency (middle), and higher resolution plot of wavelength relative to average linear rate (lower), all plotted versus time. ({\bf B}) Measurement of a broadband step-tuned laser as for laser in panel {\bf A}. Lower panel is a zoom-in to illustrate resolution of the measurement. ({\bf C}) Spectroscopy of H$^{12}$C$^{14}$N gas. A vibronic level of H$^{12}$C$^{14}$N gas at 5 Torr is resolved using the laser in panel {\bf A}. ({\bf D}) Energy level diagram showing transitions between ground state and 2$\nu_1$ levels. The measured (reference) transition wavenumbers are noted in red (blue).}
\end{center}
\label{figure2}
\end{figure*}

\begin{figure*}
\captionsetup{singlelinecheck=off, justification = RaggedRight}
\includegraphics{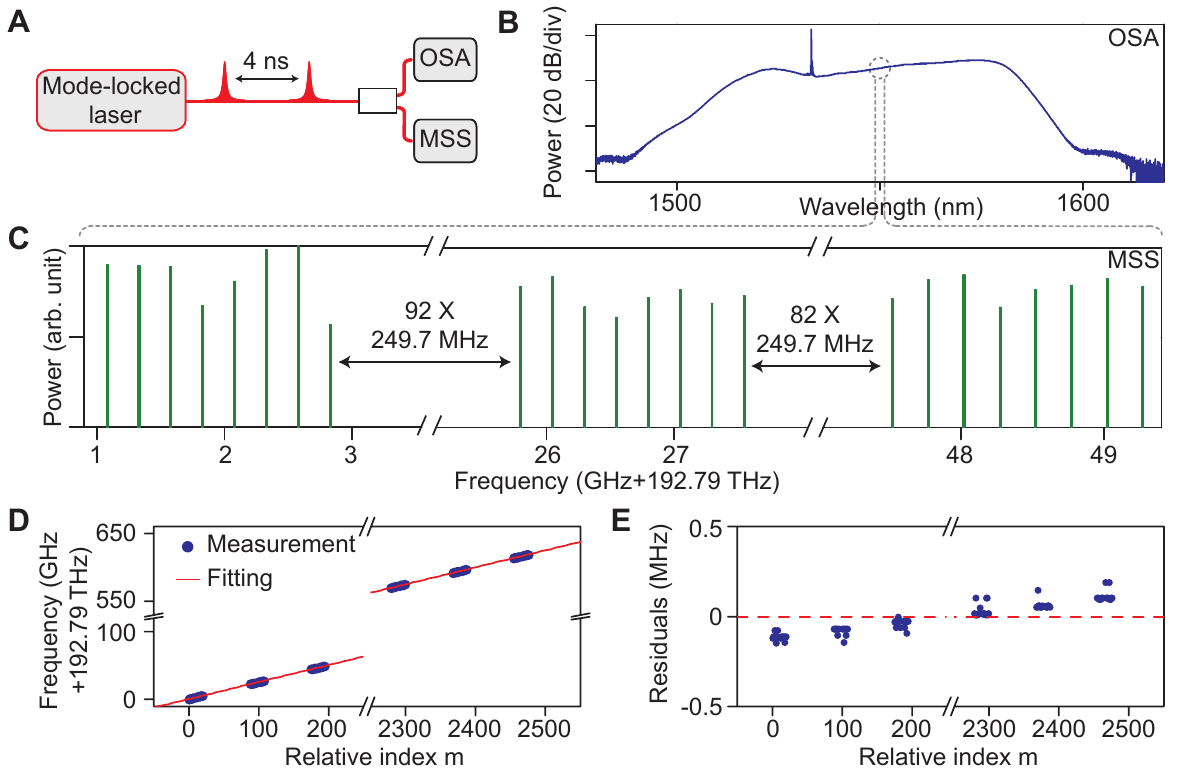}
\caption{{\bf Measurement of a fiber mode-locked laser} ({\bf A}) Pulse trains generated from a fiber mode-locked laser (FMLL) are sent into an optical spectral analyzer (OSA) and the MSS. ({\bf B}) Optical spectrum of the FMLL measured by the OSA. ({\bf C}) Optical spectrum of the FMLL measured using the MSS. Only a 60-GHz wavelength range is selected. ({\bf D}) Measured (blue) and fitted (red) FMLL mode frequencies versus index. The slope of the fitted line is set to 249.7 MHz, the measured FMLL repetition rate. ({\bf E}) Residual MSS deviation between measurement and fitted value.}
\label{figure3}
\end{figure*}

The measurement concept in the frequency domain is depicted in Fig. 1A where comb spectra from doubled-locked cw and ccw solitons are shown. The solitons are pumped from a single laser source that is modulated as shown in figure 1B to produce the two mutually-coherent pump lines at order $\mu = N$ with frequency separation $\Delta \nu$. The difference in pumping frequencies (MHz range) causes the soliton repetition rates to differ by $\Delta f_r$ which sets up a vernier effect in the respective soliton comb frequencies. As detailed elsewhere, the cw and ccw combs will experience frequency locking at order $\mu = 0$ for certain pumping frequencies \cite{yang2017counter}. This locking requires that $\Delta \nu = N \Delta f_r$. Also, because the two pump frequencies are derived from a single laser source and have a high relative frequency stability ($\Delta \nu$ is very stable), the two combs are also effectively locked at order $\mu = N$. The order $N$ can readily increased or decreased by adjusting $\Delta \nu$. The line-by-line relative frequency stability caused by this double locking is comparable to an excellent radio-frequency source.  Moreover, the frequency spacings between comb tooth pairs occur at precise integer multiples of $\Delta f_r$ (the stability of which is ensured through the relation $\Delta \nu = N \Delta f_r$), and thereby creates an extremely stable optical frequency vernier for mapping of the comb spectra into a radio frequency grid spectrum. 

The spectrometer operates as follows. A test laser frequency $f_L$ is measured using either of the following expressions: $f_L = n f_{r1,2} + \Delta f_{n1,2} + f_0$ where $n$ is the comb order nearest to the laser frequency, $f_{r1,2}$ are the comb repetition rates, $\Delta f_{n1,2}$ are the heterodyne beat frequencies of the test laser with the two frequency comb teeth at order $\mu = n$, and $f_0$ is the frequency at $\mu=0$. Comb repetition rates $f_{r1,2}$ and the beats $\Delta f_{n1,2}$ are measured by co-detection of the combs and the test laser to produce the electrical signals $V_{1,2}$ in Fig. 1B. The correlation method \cite{giorgetta2010fast} is used to determine $n$. This method can be understood as a calculation of the frequency difference $\Delta f_{n2} - \Delta f_{n1} = n \Delta f_r $ by formation of $V_1 V_2$ followed by fast Fourier transform (FFT). A typical FFT spectrum of $V_1 V_2$ is shown in Fig. 1C and gives a spectral line at $ n \Delta f_r $.  To determine $n$ requires $\Delta f_r = f_{r2}- f_{r1}$ which is measured by heterodyne of the solitons to produce electrical signal $V_3$. Figure 1D is a narrow frequency span of the FFT of $V_3$ and shows how the optical frequency vernier is mapped into a stable radio-frequency grid with line spacing $\Delta f_r $. The order corresponding to the FFT of the $V_1 V_2$ signal (Fig. 1C spectrum) is also indicated. These steps are performed automatically to provide a real time measurement of $f_L$ relative to $f_0$. To determine $f_0$ the order of a comb tooth nearest a reference laser (with known and stable frequency) is determined. This can be done, for example, by application of the correlation procedure to the reference laser. Then, as illustrated in Fig. 1B, the beat of the reference laser with this comb order is monitored for real time measurement of $f_0$ during operation of the MSS. In the current system the reference laser is stabilized using an internal molecular reference. 


The counter-propagating solitons are generated in a high-$Q$ silica microresonator with 3 mm diameter and corresponding 22 GHz soliton repetition rate \cite{lee2012chemically}. Details of the soliton generation process can be found elsewhere \cite{yi2015soliton,yi2016active,yang2017counter}. Typical optical spectra of cw and ccw solitons are plotted in Fig. 1E and span the telecommunication C-band. The distinct pumping frequencies enable repetition rate tuning to control $\Delta f_r$ through the Raman self-frequency shift \cite{milian2015solitons,karpov2016raman,yi2016theory,yang2016spatial,yang2017counter}. For example, a repetition rate difference of $\Delta f_r = 52$ kHz as seen in Fig. 1D results from a pumping frequency difference of $\Delta \nu = 4.000$ MHz ($N = 77$).



As a preliminary test, the frequency of an external-cavity-diode-laser is measured and compared against a wavemeter. Fig. 1C and 1D ($n=54$) are from this measurement. The real-time measured wavelength of the laser is presented in Fig. 1F and fluctuates within $\pm 0.02$ pm over a 5 ms time interval. The measurement is repeated from 1545 to 1560 nm and the acquired wavelengths are plotted in Fig. 1G. The data show residual deviations less than 0.1 pm versus a wavemeter measurement, which is believed to be limited primarily by the wavemeter resolution ($\pm 0.1$ pm). The systematic uncertainty of the absolute wavelength measurement in the current setup is around $\pm 4$ MHz ($\pm 0.03$ pm) and is dominated by stability of the reference laser.


The large, microwave-rate, free-spectral range of the MSS enables tracking of fast-chirping lasers in real time and discontinuous broadband tuning. Although correlation is performed with a time interval $T_W = 1/\Delta f_r$, the instantaneous frequency of the laser relative to the combs can be acquired at a much faster rate set by the desired time-bandwidth-limited resolution. To avoid aliasing of correlation measurement (i.e., to determine $n$ uniquely), the amount of frequency-chirping should not exceed the repetition rate $f_r$ within the measurement window $T_W$, which imposes a maximum resolvable chirping-rate of $ f_r\times\Delta f_r$. This theoretical limit is 1 PHz/s for the MSS and represents a boost of $100\times$ compared with previous Vernier spectrometers \cite{giorgetta2010fast}. 

To test the MSS dynamically, it is first used to measure rapid continuous-tuning of an external cavity diode laser. As shown in the upper panel of Fig. 2A, the correlation measurement evolves as the laser is tuned over multiple FSRs of the comb and thereby determines the index $n$ as a function of time. The frequency of the scanning laser is displayed at low resolution in the middle panel of Fig. 2A and shows a linear chirping-rate of $-12.4$ THz/s. Finally, the lower panel in Fig. 2A shows the measured frequency versus time at higher resolution by removing the average linear frequency ramp. As discussed in the Methods Section, the discontinuities in the measurement are caused by electrical frequency dividers used to reduce the detected signal frequency for processing by a low-bandwidth oscilloscope.  These dividers can be eliminated by using a faster oscilloscope. In Fig. 2B the MSS is used to resolve
broadband step tuning (mode hopping) of an integrated ring resonator based tunable III-V/Silicon laser diode \cite{Tran2018}.  Fast step tuning between 1551.427 nm and 1557.613 nm every 1 ms with the corresponding index $n$ stepping between $n=64$ and $n=29$ is observed. The lower panel in Fig. 2B gives a higher resolution zoom-in of one of the step regions. The data points in these measurements are each acquired over 1$\mu$s so the resolution is approximately 1 MHz.

This combination of speed and precision is also useful for spectroscopic measurements of gas-phase chemicals using tunable, single-frequency lasers. Figure 2C is an absorption line of H$^{12}$C$^{14}$N at 5 Torr obtained by a scanning laser calibrated by the MSS. The linewidth is around 2.6 GHz and the absorbance is as weak as 0.12 dB. Separate measurements on vibronic transitions between the ground state and 2$\nu_1$ states were performed. Fig. 2D summarizes the corresponding pseudo-Voigt fitting for the transition wavenumbers, which are in excellent agreement with the HITRAN database \cite{gordon2017hitran2016}.

To illustrate a measurement of more complex multi-line spectra, a fiber mode-locked laser (FMLL) is characterized as shown in Fig. 3A. For this measurement, the FMLL was first sent through a bandpass filter to prevent detector saturation. Also, the frequency extraction procedure differs and is modified to enable unique identification of many frequencies (see Supplement). The FMLL line spacing of 249.7 MHz (measured by photodetection) is not resolved in the Fig. 3B spectrum measured using a grating spectrometer. On the other hand, the reconstructed FMLL spectrum measured using the MSS is plotted in Fig. 3C; here, the comb lines are resolved and their frequency separations closely match the value measured by photo detection.  Further details on this measurement are provided in the Supplemental section. In a second study of the FMLL, the MSS is used to measure 6 closely-spaced-in-frequency groups of lines located at various spectral locations spanning 2500 free-spectral-range's of the mode locked laser. The measured frequencies are plotted in Fig. 3D. A linear fitting defined as $f_m=f_o+m f_{\mathrm{rep}} $ is plotted for comparison by using the measured FMLL repetition rate $f_{\mathrm{rep}}=249.7$ MHz where $m$ and $f_o$ represents the relative comb index and fitted offset frequency at $m=0$, respectively. The residual deviation between the measurement and linear fitting is shown in Fig. 3E and gives excellent agreement. The slight tilt observed in Fig. 3E is believed to be related to drifting of soliton repetition rates which were not monitored real-time. Also, variance of residuals within each group comes from the 300 kHz linewidth of each FMLL line. Drifting of the reference laser and FMLL carrier-envelope offset also contributes to the observed residuals across different measurements.

In conclusion, a soliton spectrometer has been demonstrated using dual-locked counter-propagating soliton microcombs. The device provides high resolution measurement of rapid continuously and step tuned lasers as well as complex multi-line spectra. In combination with a tunable laser, precise measurement of absorption spectra including random spectral access (as opposed to only continuous spectral scanning) can be performed.  Further optimization of this system could include generation of solitons from distinct mode families thereby allowing tens-of-MegaHertz repetition rate offset to be possible \cite{lucas2018spatial}. If such solitons can be dual-locked, the increased acquisition speed would enable measurement of chirping-rates close to 1 EHz/s. Operation beyond the telecommunications band would also clearly be useful and could employ soliton broadening either internally \cite{brasch2016photonic} or using on-chip broadeners \cite{lamb2018optical}. Besides the performance enhancement realized with the soliton microcombs, the use of dual-locked counter-propagating solitons provides a considerable technical simplification by eliminating the need for a second mutually phase locked comb. Also, it is interesting to note that the counter-propagating dual-locked solitons are potentially useful in a different application wherein dual-comb down conversion is used to perform TeraHertz spectroscopy \cite{kliebisch2018unambiguous}. Finally, chip integrable versions of the current device employing silicon nitride waveguides are possible \cite{yang2018bridging}. These and other recently demonstrated compact and low-power soliton systems \cite{stern2018battery,liu2018ultralow} point towards the possibility of compact microresonator soliton spectrometers.

\bigskip

\noindent\textbf{Methods}

\begin{footnotesize}
{\noindent {\bf Experimental details.} 
The bandwidth limit of the oscilloscope used in this experiment is 2.5 GHz and in order to measure frequencies $\Delta f_{n1,2}$ up to 11 GHz, microwave frequency dividers were used that function between 0.5 GHz to 10 GHz and provide an 8$\times$ division ratio. The use of these dividers created 3 GHz frequency unresolvable bands within one FSR of the optical combs, which caused the discontinuities in the lower panel in Fig. 2A.  Meanwhile, the repetition rate difference corresponding to the divided signals will also decrease proportionally by a factor of 8, which in turn reduces the maximum resolvable chirping rate to 125 THz/s. The dividers can be omitted by using a higher-bandwidth oscilloscope, which eliminates the above unresolvable bands and allows chirp-rate measurements approaching the theoretical limit.}

The pump is a fiber laser with free-running linewidth less than 2 kHz over 100 ms \cite{Lee2014spiral}. The long term stability of the soliton is maintained by introducing a feed back loop control \cite{yi2015soliton,yi2016active}.


\end{footnotesize}

\bigskip

\noindent\textbf{Acknowledgment}

\noindent The authors gratefully acknowledge the Defense Advanced Research Projects Agency (DARPA) under the SCOUT (W911NF-16-1-0548) and 
DODOS (HR0011-15-C-055) programs; the Air Force Office of Scientific Research (FA9550-18-1-0353) and the Kavli Nanoscience Institute.

\bibliography{ref}
\end{document}


\title{Supplementary Information to ``Vernier spectrometer using counter-propagating soliton microcombs''}

\author{Qi-Fan Yang$^{1,\ast}$, Boqiang Shen$^{1,\ast}$, Heming Wang$^{1,\ast}$, Minh Tran$^{2}$, Zhewei Zhang$^{1}$, Ki Youl Yang$^{1}$, Lue Wu$^{1}$, Chengying Bao$^{1}$, John Bowers$^{2}$, Amnon Yariv$^{1}$ and Kerry Vahala$^{1,\dagger}$\\
$^{1}$T. J. Watson Laboratory of Applied Physics, California Institute of Technology, Pasadena, California 91125, USA.\\
$^{2}$University of California, Santa Barbara, Department of Electrical and Computer Engineering, Santa Barbara, CA 93106, USA.\\
$^{\ast}$These authors contributed equally to this work.\\
$^{\dagger}$Corresponding author: vahala@caltech.edu}

\date{\today}

\maketitle
\nopagebreak

\section{Sample preparation and soliton generation}

\begin{figure*}
\captionsetup{singlelinecheck=off, justification = RaggedRight}
\begin{center}
\includegraphics{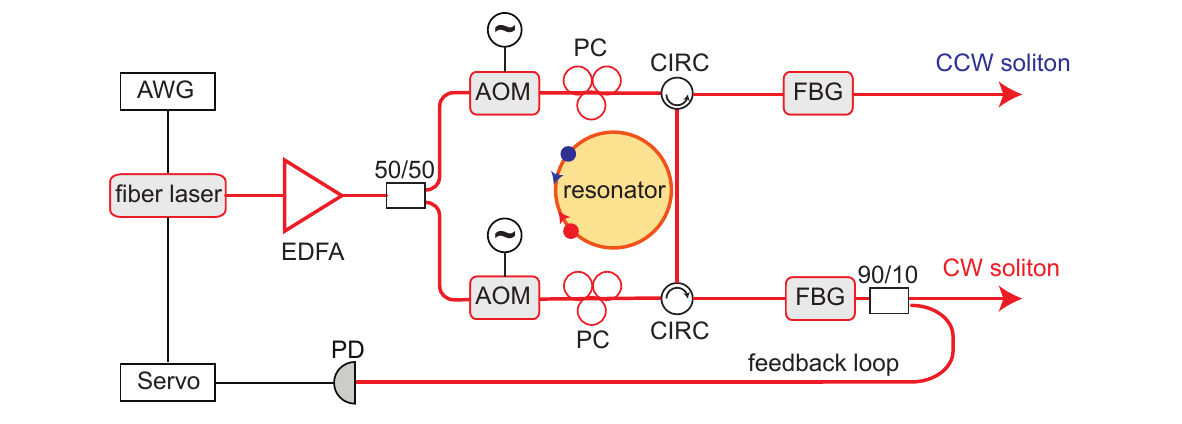}
\caption{
\label{S1}
{\bf Detailed experimental setup for soliton generation.} AWG: arbitrary waveform generator; EDFA: erbium-doped fiber amplifier; AOM: acousto-optic modulator; PC: polarization controller; CIRC: circulator; FBG: fiber Bragg grating; PD: photodetector.}
\end{center}
\end{figure*}

The silica microresonators are fabricated on a 4-inch silicon wafer with a 8-µm-thick thermally-grown silica layer. The detailed fabrication process can be found elsewhere \cite{lee2012chemically}. The intrinsic quality factor of the resonators used in this work ranges between 200 to 300 million. Light is coupled to the resonator via a tapered fiber; however, it is also possible to use silica resonators having an integrated silicon nitride waveguide \cite{yang2018bridging}.


The detailed experimental setup for soliton generation is illustrated in Fig. \ref{S1}. A continuous-wave fiber laser is amplified by an erbium-doped fiber amplifier (EDFA), and split by a 50/50 directional coupler for clockwise (cw) and counter-clockwise (ccw) soliton generation. Two acousto-optic modulators (AOMs) are used to independently control the pump frequency and power in both directions. The pump power in each direction is around 200 mW and is attenuated after the resonator by a fiber Bragg grating (FBG). The filtered transmitted power for the cw direction is split by a 90/10 directional coupler and the 10 percent output port is used in a servo control loop to stabilize the solitons. By scanning the laser from the blue side to the red side of the resonance solitons form simultaneously in both directions with characteristic ``step-like'' features in the transmitted power scan \cite{herr2014temporal,yi2015soliton,yang2017counter}. A fast power modulation is first applied to extend the existence range of the solitons. This is followed by activation of a servo control loop to stabilize the solitons at a selected power by feedback to the pump laser frequency \cite{yi2015soliton, yi2016active}. Using this approach, the solitons in both directions can exist indefinitely. 


\section{Characterization of soliton phase locking}

\begin{figure}
\captionsetup{singlelinecheck=off, justification = RaggedRight}
\includegraphics[width=17cm]{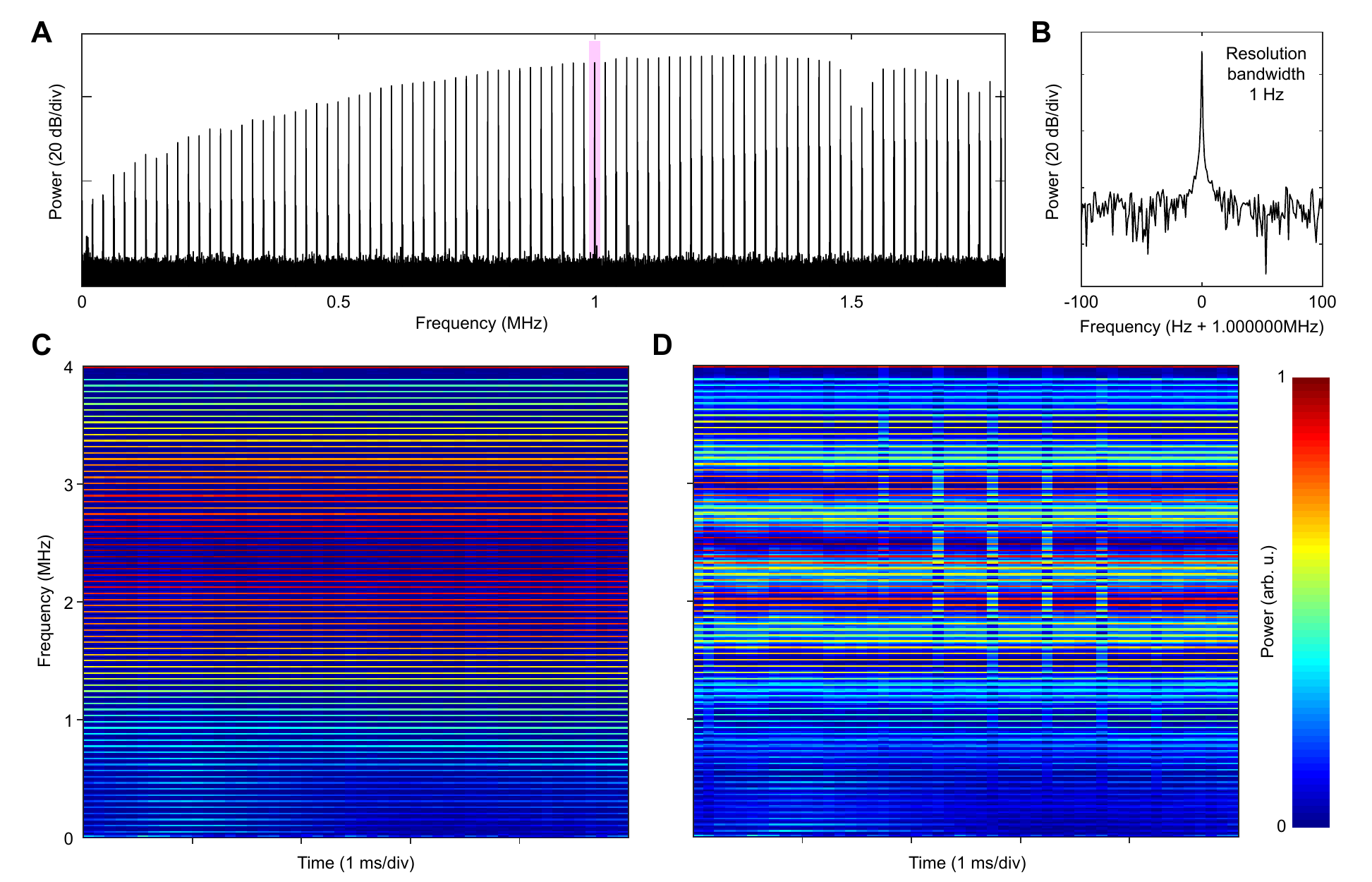}
\caption{
\label{S2}
{\bf Interferograms of cw and ccw solitons.} ({\bf A}) A typical inteferogram in frequency domain. ({\bf B}) A zoom-in of line $48$ centered at $1.000000\ \mathrm{MHz}$ (shaded region in panel A).
({\bf C}) Spectrogram of Fig. 1d in the main text showing more lines. ({\bf D}) Same as in ({\bf C}) but the frequency spacing has been deliberately chosen so that artifacts appear in the scan (see discussion in section IIIA).}
\end{figure}

The underlying mechanism leading to phase locking of counter-propagating (CP) solitons has been described in detail elsewhere \cite{yang2017counter}. Once the solitons are phase-locked, their relative frequency becomes very stable and their baseband inteferogram features sharp spectral lines (linewidths well below 1 Hz) in the frequency domain (Fig. \ref{S2}A and \ref{S2}B). To ensure that the solitons are locked during the measurement, the spectrogram of the CP soliton beatnotes is monitored as shown in Fig. 1D in the main text. In the locked case, sharp, horizontal spectral lines persist over the measurement time (Fig. \ref{S2}C).
The acquisition time window $T_W$ should be chosen to be integer multiples of $1/\Delta f_r$, where $\Delta f_r$ is the repetition rate difference, so that the frequency of the beatnotes can be accurately resolved (details can be found in section IIIA). If this is not the case, artifacts will appear in the spectrogram due to misalignment of the frequency grids (Fig. \ref{S2}D).

\section{Signal processing}

\subsection{General processing algorithm}
Through heterodyne of the test laser with the nearest comb teeth, the phase $\psi$ of the test laser is related to the electrical signals $V_{1,2}$ by
\begin{equation}
V_{1,2}\propto \cos(\psi - 2\pi \nu_{n1,2}t),
\end{equation}
where $\nu_{n1,2}$ represent the frequencies of nearest comb teeth and have order $n$. We also have $\nu_{n2}-\nu_{n1}=n\Delta f_r$ as a result of the CP soliton locking. A Hilbert transform is used to extract the time-dependent phase $\psi -2\pi \nu_{n1,2} t$ from $ V_{1,2} $ which thereby gives the heterodyne frequencies via
\begin{equation}
\Delta f_{n1,2}=\dot{\psi}/2\pi-\nu_{n1,2},
\end{equation}
Each data point of $\Delta f_{n1,2}$ is obtained by linear fitting of the phase over a specified time interval that sets the frequency resolution. Similarly, the heterodyne frequency between the reference laser and the soliton comb can be retrieved to determine the frequency $f_0$ (see discussion in main text).

The Fourier transform of the product $V_1V_2$ is given by
\begin{equation}
\begin{split}
\widetilde{V_1V_2}(f)&\propto \int_0^{T_W} \frac{e^{i(\psi-2\pi \nu_{n1}t)}+e^{-i(\psi-2\pi \nu_{n1}t)}}{2} \frac{e^{i(\psi-2\pi \nu_{n2}t)}+e^{-i(\psi-2\pi \nu_{n2}t)}}{2} e^{-2\pi i f t} \mathrm{d}t\\
&\propto \delta(|f|-n\Delta f_r),
\end{split}
\end{equation}
where sum frequency terms in the integrand are assumed to be filtered out and are therefore discarded. To accurately extract the above spectral signal the acquisition time window $T_W$ should be an integer multiple of $1/\Delta f_r$, which is also related to the pump frequency offset $\Delta\nu$ by $T_W=N_WN/\Delta\nu$ where $N$ is the pump order and $N_W$ is an integer. Moreover, the number of sampled points, which equals the product of oscilloscope sampling rate $f_\mathrm{samp}$ and $T_W$, should also be an integer (i.e., $f_\mathrm{samp} N_WN/ \Delta\nu$ is an integer). In this work, $f_\mathrm{samp}$ is usually set to 2.5 or 5 GHz/s and it is found that simple adjustment of $\Delta\nu$ is sufficient to satisfy this condition.
As a result it is not necessary to synchronize the oscilloscope to external sources. It is noted that this method is simpler than the asynchronous detection used in previous work \cite{giorgetta2010fast}.

 On account of the limited bandwidth of the oscilloscope used in work, it was necessary to apply electrical frequency division to the detected signals for proceesing by the oscilloscope. When frequency dividers are used (division ratio $r=8$), the divided electrical signals (indicated by superscript d) yield
\begin{equation}
V_{1,2}^{\mathrm{d}}\propto \cos((\psi - 2\pi \nu_{n1,2}t)/r).
\end{equation}
As a result, the divided frequencies also satisfy $\Delta f_{n1,2}^{\mathrm{d}}=\Delta f_{n1,2}/r$ and the correlation between the divided signals scales proportionally by
\begin{equation}
\Delta f_{n1}^{\mathrm{d}}-\Delta f_{n2}^{\mathrm{d}}=n\Delta f_r/r.
\end{equation}
Therefore the required resolution bandwidth to resolve the ambiguity $n$ from the measured correlation is $\Delta f_r / r$ which increases the minimal acquisition time to $T_W^{\mathrm{d}}=r T_W$.

\subsection{Absorption spectroscopy}
To perform the absorption spectroscopy the laser transmission through the H$^{12}$C$^{14}$N gas cell is recorded while the laser is continuously scanning. A portion of the laser signal is also meaured in the MSS to determine its frequency during the scan. A pseudo-Voigt lineshape (linear combination of Gaussian and Lorentz profile) is fitted to the spectrum and the central frequency is then extracted.

\begin{figure}
\captionsetup{singlelinecheck=off, justification = RaggedRight}
\includegraphics[width=17cm]{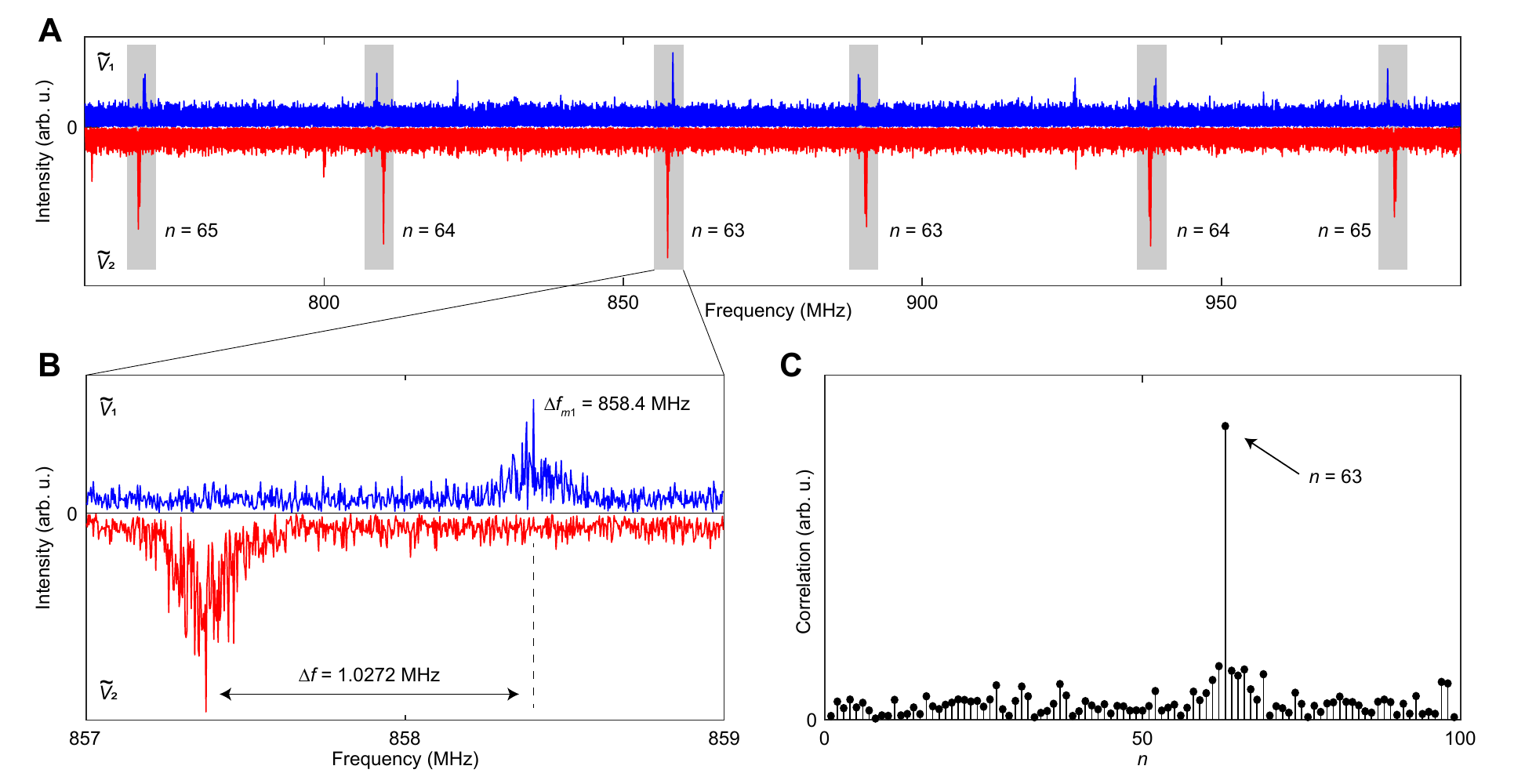}
\caption{
\label{S3}
{\bf Multi-frequency measurements.} {\bf (A)} A section of $\tilde{V}_{1,2}$. Pairs of beatnotes coming from the same laser are highlighted and the derived $n$ value is marked next to each pair of beatnotes. {\bf (B)} Zoom-in on the highlighted region near 858 MHz in {\bf (A)}. Two beatnotes are separated by $1.0272$ MHz. {\bf (C)} Cross-correlation of $\tilde{V}_{1}$ and $\tilde{V}_{2}$ is calculated for each $n$ and the maximum can be found at $n=63$.}
\end{figure}

\subsection{Mode-locked laser measurement}
The algorithm used here to extract a large number of frequencies simultaneously using the MSS is different from the previous single-frequency measurements. Rather than multiplying the signals $V_1$ and $V_2$ followed by Fast Fourier Transform (FFT) in order to determine the microcomb order, we directly FFT the signals $V_1$ and $V_2$ followed by filtering and then frequency correlation. This avoids the generation of ambiguities. To explain the approach, first consider an implementation similar to that reported in the main text. There, a fiber mode locked laser (FMLL) comb with free-spectral-range (FSR) of about 250 MHz was optically filtered to create a narrower frequency range of FMLL laser lines extending over only a few microcomb lines. The signals $V_1$ and $V_2$ upon FFT therefore produce a large set of frequencies representing the individual beats of each FMLL laser line (index $m$) with microcomb modes (index $n$). Fig. S3A gives a narrow frequency span of a typical FFT generated this way for both the $V_1$ and $V_2$ signals. A zoom-in of one pair of $V_1$ and $V_2$ signals is provided in Figure S3B and a remarkably precise frequency separation between the beats (in view of the spectral breadth of each beat) can be determined by correlating the upper (blue) and lower (red) spectrum (see Fig. S3C). This precision results from the underlying high relative frequency stability of the cw and ccw microcomb frequencies. As described in the main text this frequency separation is a multiple of $\Delta f_r$ and plot of the correlation versus the frequency separation (in units of $\Delta f_r$) is provided in Fig. S3C where the peak of the correlation gives the index $n=63$ for this pair of beat frequencies. Proceeding this way for each pair of peaks in Fig. S3A allows determination of $n$ from which the frequency of the corresponding FMLL line can be determined. It is interesting to note that in Fig. 3A, there are two sets of peaks that give $n$=63, 64 and 65. These correspond to FMLL lines that are higher and lower in frequency relative to the microcomb modes with indices $n$=63, 64 and 65. The relative alignment of the blue and red peaks which switches sign for these sets of beat frequencies allows determination of which FMLL line is lower and higher in frequency relative to the microcomb lines.


To provide more rigor to this explanation, the electrical signals consist of multiple beat components components given by,
\begin{equation}
V_{1,2}=\sum_mV_{m1,2},\ \ V_{m1,2}\propto\cos(\psi_m-2\pi \nu_{\mu(m)1,2}t),
\end{equation}
where $\psi_m$ and $\nu_{\mu(m)1,2}$ represent the phase of the $m$-th FMLL mode and the frequencies of the microcomb order nearest to this FMLL mode, respectively, where $\mu(m)$ denotes the comb order nearest the $m$-th FMLL mode. As described in the main text the frequencies $\nu_{\mu(m)1,2}$ are related to the repetition rate difference by $\nu_{\mu(m)2}-\nu_{\mu(m)1}=\mu(m)\Delta f_r$. The FFT of $V_{1,2}$ is denoted by $\tilde{V}_{1,2}$ and the correlation given in Fig. S3C (and used to determine the comb order $n$ of each spectral component) is given by,
\begin{equation}
\begin{split}
\int_{\Delta f_{m1}-\kappa/2}^{\Delta f_{m1}+\kappa/2}\tilde{V}_{1}(f)\tilde{V}_{2}^{*}(f+n\Delta f_r) \mathrm{d}f&\approx\int_{-\infty}^{\infty} \mathrm{d}f\int V_{m1}(t) e^{2\pi ift} \mathrm{d}t\int V_{m2}(t') e^{-2\pi i(f+n\Delta f_r)t'} \mathrm{d}t'\\
&=\int V_{m1}(t)V_{m2}(t) e^{-2\pi i n \Delta f_r t}\mathrm{d}t\\
&\propto\int \frac{e^{i(\psi_m-2\pi \nu_{\mu 1}t)}+e^{-i(\psi_m-2\pi \nu_{\mu 1}t)}}{2} \frac{e^{i(\psi_m-2\pi \nu_{\mu 2}t)}+e^{-i(\psi_m-2\pi \nu_{\mu 2}t)}}{2} e^{-2\pi i n \Delta f_r t} \mathrm{d}t\\
&\propto\delta(\mu(m)-n).
\end{split}
\end{equation}
where $\Delta f_{m1}$ denotes the peak frequency of the beatnote, $\kappa$ is a predetermined range of integration to cover the linewidth of the beatnote (here $\kappa = 2$ MHz), and where sum frequency terms in the integrand have been discarded. Therefore for each spectral component $m$, its associated microcomb order number $\mu(m)$ can be determined by varying $n$ in the above correlation until it reaches maximum (see Fig. S3C). The $n$ value with the maximum correlation will be assigned to the peak as the tooth number $\mu(m)$ and then the absolute frequency can be recovered.


The limit of this process to accomodate more FMLL frequencies is much higher than that given by the filter bandwidth studied in this work. It is instead set by the spectral density of FMLL-microcomb beat lines that can be reasonably resolved within the microcomb FSR spectral span.  


\bibliography{ref}